\documentclass[usenatbib]{mn2e}
\usepackage{epsfig}

\title{Photometric Properties of Lyman-break Galaxies at $z=3$ 
in Cosmological SPH Simulations}

\author[Nagamine, Springel, Hernquist, \& Machacek]
  {K.~Nagamine$^1$\thanks{Email: knagamin@cfa.harvard.edu},
  V.~Springel$^2$\thanks{Email: volker@mpa-garching.mpg.de},
  L.~Hernquist$^1$\thanks{Email: lars@cfa.harvard.edu},
  M.~Machacek$^1$\thanks{Email:
  mmachacek@cfa.harvard.edu}\vspace{0.3cm}\\ $^1$Harvard-Smithsonian
  Center for Astrophysics, 60 Garden Street, Cambridge, MA 02138,
  U.S.A. \\ $^2$Max-Planck-Institut f\"{u}r Astrophysik,
  Karl-Schwarzschild-Stra\ss{}e 1, 85740 Garching bei M\"{u}nchen,
  Germany}

\pagerange{\pageref{firstpage}--\pageref{lastpage}}
\pubyear{2003}

\newcommand {\Mstar}{M_{\rm star}}
\newcommand{\U}{U_n}

\newcommand{\Lam}{\Lambda}

\newcommand{\yr}{{\rm yr}}
\newcommand{\mpc}{{\rm Mpc}}
\newcommand{\kpc}{{\rm kpc}}

\newcommand{\Msun}{\rm M_{\odot}}

\newcommand{\hinv}{h^{-1}}
\newcommand{\himpc}{\hinv{\rm\,Mpc}}
\newcommand{\hikpc}{\hinv{\rm\,kpc}}
\newcommand{\himsun}{\,\hinv{\Msun}}

\newcommand{\Om}{\Omega_{\rm m}}
\newcommand{\Ol}{\Omega_{\Lam}}
\newcommand{\Ob}{\Omega_{\rm b}}

\newcommand{\Lbox}{{L_{\rm box}}}

\newcommand{\etal}{et~al.}

\newcommand{\beq}{\begin{eqnarray}}
\newcommand{\eeq}{\end{eqnarray}}

\begin{document}

\maketitle

\label{firstpage}


\begin{abstract}

We study the photometric properties of Lyman-break
galaxies (LBGs) formed by redshift $z=3$ in a set of large
cosmological smoothed-particle hydrodynamics (SPH) simulations of the
$\Lambda$ cold dark matter ($\Lambda$CDM) model. Our numerical
simulations include radiative cooling and heating with a uniform UV
background, star formation, supernova feedback, and a phenomenological
model for galactic winds.  Analysing a series of
simulations of varying box size and particle number allows us to
isolate the impact of numerical resolution on our results.  We compute
spectra of simulated galaxies using a population synthesis model, and
derive colours and luminosity functions of galaxies at $z=3$ after
applying local dust extinction and absorption by the intergalactic
medium (IGM). We find that the simulated galaxies have $\U-G$ and
$G-R$ colours consistent with observations, provided that intervening
absorption by the IGM is applied.  The observed properties of LBGs,
including their number density, colours, and luminosity functions, can
be explained if LBGs are identified with the most massive galaxies at
$z=3$, having typical stellar mass of $M_{\star} \sim 10^{10}\himsun$, a
conclusion broadly consistent with earlier studies based on
hydrodynamic simulations of the $\Lam$CDM model. We also find that
most simulated LBGs were continuously forming stars at a high rate for
more than one Gyr up until $z=3$, but with numerous starbursts lying on 
top of the continuous component. Interestingly, our simulations
suggest that more than 50\% of the total stellar mass and star
formation rate in the Universe are accounted for by galaxies that are
not detected in the current generation of LBG surveys.
\end{abstract}



\section{Introduction}
\label{section:intro}

Based on surveys made with modern large telescopes, the number of 
observed Lyman-break galaxies \citep[LBGs; e.g.][]{Steidel95} has 
reached about $\sim 1000$ \citep{Ste03} by now, providing a unique 
window for studying early galaxy formation.  
In these observations, a large sample of
LBG-candidates is first selected as ``drop-outs'' in certain sets of
colour-colour planes (e.g. $\U-G$ vs. $G-R$), and then later
spectroscopic follow-up observations are performed to obtain the
redshifts of the identified candidates. Most surveys to date have been
carried out at optical wavelengths, corresponding to the rest-frame
far ultra-violet (UV) at redshift $z=3$. The LBGs found in this way
show strong clustering \citep{Ade98, Giavalisco98, Steidel98}, which
has generally been interpreted as indirect evidence that LBGs reside
in massive dark matter halos. Several studies of LBGs based on
semi-analytic models for galaxy formation agree with the massive dark
matter halo hypothesis \citep[e.g.][]{Mo96, Bau98, Kau99, Mo99}, but a
numerical study by \citet{Jing} using collisionless N-body simulations
expressed some concerns whether the clustering of dark matter halos
of $<10^{12}\himsun$ can explain the observed clustering strength of
LBGs. However, a later 
investigation by \citet{KHW99} argued that the CDM
models have no difficulty in explaining the strong observed clustering
of LBGs, although the substantial box size effects in their analysis
introduced some uncertainty.

More recently, LBGs have also been studied in the near infrared (IR)
\citep*{Sawicki98, Papovich01, Pettini01, Sha01, Rudnick01}.  Near IR
observations are less affected by dust obscuration and directly probe
the rest-frame optical properties of LBGs at $z=3$.  Therefore, they
allow a more reliable derivation of the stellar masses of LBGs.  Based
on these observations, it has been suggested that the value of
the extinction lies in the range $E(B-V)\approx 0.0-0.3$ with a median 
value of 0.15, and that a
significant old stellar population exists in LBGs at $z=3$, with a
stellar mass of $\approx 10^{10}\Msun$ \citep{Papovich01, Sha01}.  The
existence of such a stellar component would again suggest that LBGs
are embedded in massive dark matter halos which have continuously
formed stars over an extended period of roughly one Gyr up to
$z=3$. These systems would then most likely evolve into elliptical
galaxies at the present day, or into the spheroidal components of
massive spiral galaxies.

However, in a competing model, LBGs have been suggested to be
merger-induced starbursting systems associated with low-mass halos
\citep*{Lowenthal97, Sawicki98, Som01}, and in some cases
merger-induced starbursts are given a crucial role even when LBGs are
the most massive galaxies at their time \citep{Som01}. Given that the
merger rate is expected to be quite high at $z\simeq 3$, these
scenarios provide an interesting alternative to the more conventional
picture which associates LBGs with the most massive systems.

Self-consistent hydrodynamic simulations are an ideal tool for trying
to distinguish between these different scenarios for the nature of
LBGs.  \citet{Dave99} and \citet{Wei02} were the first to employ
smoothed particle hydrodynamics (SPH) simulations to this end.
However, their box size of $11.1\,\himpc$ represented an important
limitation, because the space-density of LBGs is so low that only a
few of them can be found in a volume of this size, as we will discuss
further in Section~\ref{section:color-color}. Therefore, simulations
with a larger box size are desirable to obtain larger samples of
simulated LBGs.  \citet{Nag02} used an Eulerian hydrodynamic
simulation with a box size of $\Lbox=25\,\himpc$, tracing the merger
history of galaxies from $z=5$ to $z=0$. The results of these earlier
numerical studies were consistent with each other, and agreed
reasonably well with the observations, within the uncertainties. In
particular, the median stellar masses of LBGs were predicted to be
$\sim 10^{10}\himsun$, and the simulated galaxies were experiencing
significant star formation rates ($>30\,\Msun/\yr$) for extended
periods of time ($\ge 1$ Gyr).

\begin{table*}
\begin{center}
\begin{tabular}{cccccccc}
\hline
Run & Box Size & ${N_{\rm p}}$ & $m_{\rm DM}$ & $m_{\rm gas}$ &
$\epsilon$ & $z_{\rm end}$ & wind \\
\hline
\hline
O3  & 10.00 & $2\times 144^3$ &  $2.42\times 10^7$ & $3.72\times 10^6$ &2.78 & 2.75 & none \cr
P3  & 10.00 & $2\times 144^3$ &  $2.42\times 10^7$ & $3.72\times 10^6$ &2.78 & 2.75 & weak \cr
Q3  & 10.00 & $2\times 144^3$ &  $2.42\times 10^7$ & $3.72\times 10^6$ &2.78 & 2.75 & strong \cr
Q4  & 10.00 & $2\times 216^3$ &  $7.16\times 10^6$ & $1.10\times 10^6$ &1.85 & 2.75 & strong \cr
Q5  & 10.00 & $2\times 324^3$ &  $2.12\times 10^6$ & $3.26\times 10^5$ &1.23 & 2.75 & strong \cr
\hline                                                                  
D4  & 33.75 & $2\times 216^3$ &  $2.75\times 10^8$ & $4.24\times 10^7$ &6.25 & 1.00 & strong \cr
D5  & 33.75 & $2\times 324^3$ &  $8.15\times 10^7$ & $1.26\times 10^7$ &4.17 & 1.00 & strong \cr
\hline                                                                  
G5  & 100.0 & $2\times 324^3$ &  $2.12\times 10^9$ & $3.26\times 10^8$ &8.00 & 0.00 & strong \cr
G6  & 100.0 & $2\times 486^3$ &  $6.29\times 10^8$ & $9.67\times 10^7$ &5.00 & 0.00 & strong \cr
\hline
\end{tabular}
\caption{Simulations employed in this study.  The box size is given in
units of $\himpc$, ${N_{\rm p}}$ is the particle number of dark matter
and gas (hence $\times\, 2$), $m_{\rm DM}$ and $m_{\rm gas}$ are the
masses of dark matter and gas particles in units of $\himsun$,
respectively, $\epsilon$ is the comoving gravitational softening
length in units of $\hikpc$, and $z_{\rm end}$ is the ending redshift
of the simulation. The value of $\epsilon$ is a measure of spatial
resolution.  From the top to the bottom row, we collectively call the
first 5 runs (O3 to Q5) `Q-series', D4 \& D5 are called `D-series',
and G5 \& G6 are called `G-series'.
\label{table:sim}}
\end{center}
\end{table*}

In this paper, we improve on the earlier numerical studies of LBGs by
using a new set of high-resolution numerical simulations. These
simulations are based on a novel model for the physics of star
formation and feedback, and they use a more accurate implementation of
SPH. For the first time, we also systematically study the effects of
resolution and box size in the context of simulated LBG galaxies. The
treatment of star formation and feedback we use is based on a
sub-resolution multi-phase description of the dense, star-forming
interstellar medium (ISM), and a phenomenological model for strong
feedback by galactic winds, as recently proposed by
\citet{SH03a}. This model has been shown to provide converged star
formation rates for well-resolved galaxies, with a cosmic star
formation history consistent with recent observations
\citep{SH03b, Her03}.\footnote[3]{We 
note an error in figure 12 of \citet{SH03b} in which the observational
estimates of the SFR were plotted too high by a factor of
$h^{-1} = 1.4$.  When corrected, the observed points are in better agreement
with the theoretical estimates; see astro-ph/0206395.}
The inclusion of winds was motivated by the realisation
that galactic outflows at high redshift \citep{Pet02} likely play a
key role in distributing metals into the intergalactic medium
\citep[e.g.][]{Aguirre01a, Aguirre01b}, as well as being important for
the regulation of star formation activity. In fact, winds may also
alter the distribution of neutral gas around galaxies \citep{Ade03},
although the details of how this process may happen remain
unclear \citep[e.g.][]{Croft02, Kol03, Bruscoli}. 
In passing, we note that both \citet{Des} and \citet{Maselli} have 
found that the Lyman-$\alpha$ transmissivity close to LBGs, as measured
by \citet{Ade03}, is better reproduced if LBGs are identified as dwarf
starbursting galaxies as proposed in \citet{Som01} and \citet{WW}.
We will discuss the work of \citet{WW} in Section~\ref{section:discussion}.
Together with the increase in numerical resolution provided by our 
simulations, it is of interest to see how our refined physical 
modelling modifies the predictions for LBG properties within the 
$\Lambda$CDM scenario.

This paper is organised as follows. In
Section~\ref{section:simulation}, we briefly introduce the numerical
parameters of our simulation set. In Section~\ref{section:method}, we
then describe our method for computing spectra of simulated galaxies
both in the rest-frame and the observed frame.  In
Sections~\ref{section:color-color} and \ref{section:colour-magnitude},
we show the colour-colour diagrams and colour-magnitude diagrams of
simulated galaxies, and we discuss the number density of
colour-selected LBGs, as well as the stellar masses of LBGs at $z=3$.
We then investigate the rest-frame $V$-band luminosity function and
observed $R$-band luminosity function in Section~\ref{section:lf},
followed by an analysis of the star formation histories of LBGs in
Section~\ref{section:sf}. Finally, we summarise and discuss the
implications of our work in Section~\ref{section:discussion}.


\section{Simulations}
\label{section:simulation}

We analyse a large set of cosmological SPH simulations with varying
box size, mass resolution and feedback strength, as summarised in
Table \ref{table:sim}. Our box size ranges from 10 to 
$100\,\himpc$ on a side, with particle numbers between $2\times 144^3$ 
and $2\times 486^3$, giving gaseous mass resolutions in the range 
$3.3 \times 10^5$ to $3.3\times 10^8\himsun$.  
These simulations are partly taken from a study of the cosmic star 
formation history by \citet{SH03b}, supplemented by additional runs 
with weaker or no galactic winds. A similar set of simulations was
used by \citet{Nag03a, Nag03b} to study the properties of damped
Lyman-$\alpha$ absorbers, but here we analyse the `G6'-run which has 
higher resolution than the `G4'-run used in the previous studies. 
The simulations with the same box size are run with the same 
initial condition.

There are three main novel features to our simulations.  
First, we use the new ``conservative entropy'' formulation of 
SPH \citep{SH02} which explicitly conserves entropy (in regions 
without shocks), as well as momentum and energy, even when one allows 
for fully adaptive smoothing lengths \citep[see e.g. ][]{Hern93}.
This formulation moderates 
the overcooling problem present in earlier formulations of SPH 
\citep[see also][]{Yoshida02, Pearce99, Croft01}.

Second, highly over-dense gas particles are treated with an effective
sub-resolution multiphase ISM model, as described by \citet{SH03a}.  
In this model, each gas particle represents a statistical mixture of 
cold clouds and a hot ambient phase. Cold clouds can grow by radiative 
cooling out of the hot medium, and form the reservoir of neutral gas
for star formation. Once star formation takes place, supernova 
explosions deposit energy into the hot gas, heating the gas and 
evaporating the cold clouds, transferring cold gas back into the hot phase. 
This feedback establishes a self-regulated cycle of star formation.

Third, a phenomenological model for galactic winds is implemented. 
In this model, gas particles are stochastically driven out of the 
dense star-forming region by assigning an extra momentum in random 
directions, with a rate and magnitude chosen to reproduce mass-loads 
and wind speeds similar to those observed.  (See \citet{SH03a} for a 
detailed discussion of this method.)
Most of our simulations employ a ``strong'' wind of speed $484\,{\rm
km\,s^{-1}}$, but for the O3, P3, \& Q3-runs, we also varied the wind
strength to examine the effect of feedback from galactic winds.  
The runs with $10\,\himpc$ boxes are collectively called the `Q-series', 
and the resolution is increased from Q3 to Q4, and to Q5.
The other series (`D-'and `G-Series') extend the strong wind results 
to larger box sizes and hence lower redshift.  Our naming convention 
is such that runs of the same model (box size and included physics) 
are designated with the same letter, with an additional number 
specifying the particle resolution.

Our calculations include a uniform UV background radiation field with a
modified \citet{Haa96} spectrum, where reionisation takes place at
$z\simeq 6$ \citep[see][]{Dave99} as suggested by the quasar observations
\citep[e.g.][]{Beck01} and radiative transfer calculations of the
impact of the stellar sources in our simulations on the IGM
\citep[e.g.][]{Sok03}.  The early reionisation at higher redshift, 
as suggested by the WMAP satellite, should not affect the results presented 
in this paper because we are mainly dealing with halos with virial 
temperatures above $10^4$\,K and the infalling gas will radiatively 
cool even if it was photoionised by the reionisation at $z>10$ before
falling into the halo \citep[see][]{SH03b}.
The radiative cooling and heating rate is computed as described by 
\citet{Katz96} assuming that the gas is optically thin and in 
ionisation equilibrium.
The adopted cosmological parameters of all runs are 
$(\Om,\Ol,\Ob,\sigma_8, h)= (0.3, 0.7, 0.04, 0.9, 0.7)$. 
The simulations were performed on the Athlon-MP
cluster at the Center for Parallel Astrophysical Computing (CPAC) at
the Harvard-Smithsonian Center for Astrophysics, using a modified
version of the parallel {\small GADGET} code \citep{Gadget}.


\section{Method}
\label{section:method}

We identify simulated galaxies as isolated groups of stars using a
simplified variant of the {\small SUBFIND} algorithm proposed by
\citet{Spr01}. In detail, we first compute an adaptively smoothed
baryonic density field for all star and gas particles, allowing us to
robustly identify centres of individual galaxies as isolated density
peaks.  We find the full extent of these galaxies by processing the
gas and star particles in the order of declining density, adding
particles one by one to the galaxies.  If all of the 32 nearest
neighbours of a particle have lower density, this particle is considered 
to be a new galaxy `seed'. Otherwise, the particle is attached to the 
galaxy that its nearest denser neighbour already belongs to. 
If the two nearest denser neighbours belong to different galaxies,
and one of these galaxies has less than 32 particles, these galaxies
are merged. If the two nearest denser neighbours belong to different 
galaxies and both of these galaxies have more than 32 particles,
then the particle is attached to the larger group of the two, 
leaving the other one intact.
Finally, we restrict the set of gas particles processed in
this way to those particles which have at least a density of $0.01$
times the threshold density for star formation, i.e. 
$\rho_{\rm th}=8.6\times 10^6 h^2\Msun\kpc^{-3}$ \citep[see][for details 
on how this parameter is determined]{SH03a}.  Note that we are not
interested in the gaseous components of galaxies -- we only include
gas particles because they make the method more robust. Since most
galaxies contain very dense star-forming gas, such a method make it 
particularly easy to select galaxies when the gas is included.

We found that the above method robustly links up star particles that
belong to the same isolated galaxy. A simpler FoF algorithm with a
small linking length can achieve a largely similar result, but the
particular choice for the linking length one needs to make in this
method represents a problematic compromise, either leading to
artificial merging of galaxies if it is selected too large, or to loss
of star particles that went astray from the dense galactic core, if
selected too small.  Note that, unlike in the detection of dark matter
substructures, no gravitational unbinding algorithm is needed to
define the groups of stars that make up the galaxies formed in the
simulations.

We only consider galaxies with at least 32 particles (star and gas 
combined) in our subsequent analysis. Each stellar particle contained 
in them is tagged by the simulation code with its mass, formation time, 
and metallicity of the gas particle that it formed out of.  
Based on these three tags, we compute the emission from each stellar 
particle, and co-add the flux
from all particles for a given galaxy to obtain the spectrum of the
simulated galaxy. We use the population synthesis model GISSEL99
\citep{BClib} that assumes the \citet{Salpeter} initial mass function
with a mass range of $[0.1, 125]\,\Msun$.

Once the intrinsic spectrum is computed, we apply the Calzetti extinction 
law \citep{Calzetti} with three different values of $E(B-V)=0.0, 0.15, 
0.3$ to investigate the effects of extinction. These values span the 
range of $E(B-V)$ estimated from observations of LBGs at $z=3$ 
\citep{Ade00, Sha01}. Rest-frame colours and luminosity functions of the 
simulated galaxies are then derived using the spectra computed in this 
manner. 

To obtain the spectra in the observed frame, we redshift the spectra 
and apply absorption by the IGM following the prescription by 
\citet{Madau95}.  Once the redshifted spectra in the observed frame
are obtained, we convolve them with various filter functions,
including $\U, G, R$ \citep{Steidel93} and standard Johnson bands, and 
compute the magnitudes in both AB and Vega systems.  Apparent $\U$, $G$, $R$
magnitudes are computed in the AB system to compare our results with
\citet{Ste03}, while the rest-frame V-band magnitude is computed in
the Vega system to compare our results with those of \citet{Sha01}.

\begin{figure*}
\begin{center}
\epsfig{file=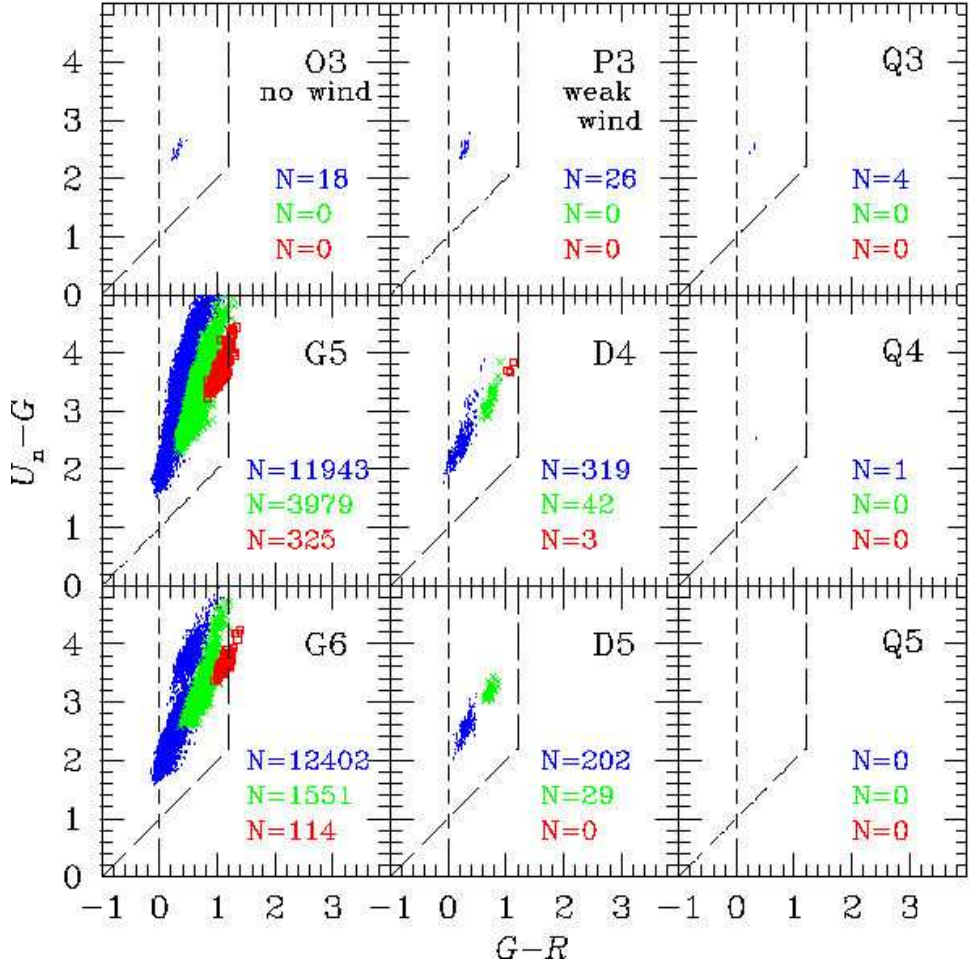,height=5in,width=5in, angle=0} 
\caption{Colour-colour diagrams ($\U-G$ vs. $G-R$) of simulated galaxies
at $z=3$ with apparent $R$ magnitudes brighter than $R<25.5$.  Three
different symbols represent three different values of extinction:
$E(B-V)=0.0$ (blue dots), 0.15 (green crosses), and 0.3 (red open
squares). The number of simulated galaxies of each type is indicated
from top to bottom in each panel, respectively.  The dashed lines
mark the colour-colour selection criteria used by \citet{Ste03} to
identify LBG candidates at $z\sim 3$.}
\label{colcol_Rcut_all.eps}
\end{center}
\end{figure*}


\section{Colour-colour diagrams}
\label{section:color-color}

In Figure~\ref{colcol_Rcut_all.eps}, we show the colour-colour
magnitude diagrams of simulated galaxies at $z=3$ in the $\U-G$
vs. $G-R$ plane, for galaxies brighter than $R=25.5$ \citep[the
magnitude cut used by][]{Ste03}.  The three different symbols
represent three different values of Calzetti extinction: $E(B-V)=0.0$
(blue dots), 0.15 (green crosses), and 0.3 (red open squares). The
long-dashed lines mark the colour-colour selection criteria applied by
\citet{Ste03} to identify LBG candidates at $z\sim 3$.

It is encouraging to see that in all the panels most of the simulated
galaxies actually satisfy the observational colour selection
criteria. This suggests that the simulated galaxies have realistic
colours compared to the observed LBGs at $z=3$. In runs with larger 
box size (D- and G-series), the distribution is wider than the one in the
$\Lbox=10\,\himpc$ runs (Q-series), and the distribution actually
 extends beyond the colour selection boundaries of \citet{Ste03}.
The box size of the Q-series ($\Lbox=10\,\himpc$) is too small to 
contain a significant sample of LBGs, and no LBGs with $R<25.5$ can
be found in the Q-series for the cases of $E(B-V)=0.15$ and 0.3.

If we had not applied the Madau (1995) absorption to the spectra, the
distribution would not have fallen into the colour selection
region. This is because the simulated galaxies would then appear too
bright in the UV, such that the distribution would fall below $\U-G <
1.0$. As the level of extinction by dust is increased from
$E(B-V)=0.0$ to 0.3, the measured points move towards the upper right
corner of each panel. This behaviour is expected for a standard
star-forming galaxy spectrum, as demonstrated in Figure 2 of
\citet{Ste03}. At the same time, the number of galaxies that satisfy
the magnitude cut of $R= 25.5$ decreases with increasing extinction,
because stronger extinction results in a redder galaxy spectrum.


Note that increasing the resolution from Q3 to Q4, and then to Q5,
also reduces the number of galaxies brighter than $R=25.5$
slightly. The same is true for D4 and D5 runs.
As the resolution is increased, not only are the most massive
halos in the simulation better resolved, but also all of their
progenitors.  The better resolution then allows a more accurate
treatment of the wind-feedback in these progenitor generations of
galaxies; the net result of this is a decrease in the final luminosity
of the brightest galaxies.  For the `Q5'-run, there are actually no
galaxies brighter than $R=25.5$.

This is not the case for the lower resolution, larger box size 
simulations where the trend appears to reverse.
In the case of the G-series, the number of galaxies brighter than 
$R=25.5$ actually increases as the resolution increases. 
As we will show in Section~\ref{section:lf_R}, this is because 
the peak of the simulated $R$-band luminosity function is still 
on the brighter side of $R=25.5$, and the increase in the number of 
galaxies near the peak of the luminosity function wins over the slight 
decrease of the number at the brightest end.

\begin{figure}
\epsfig{file=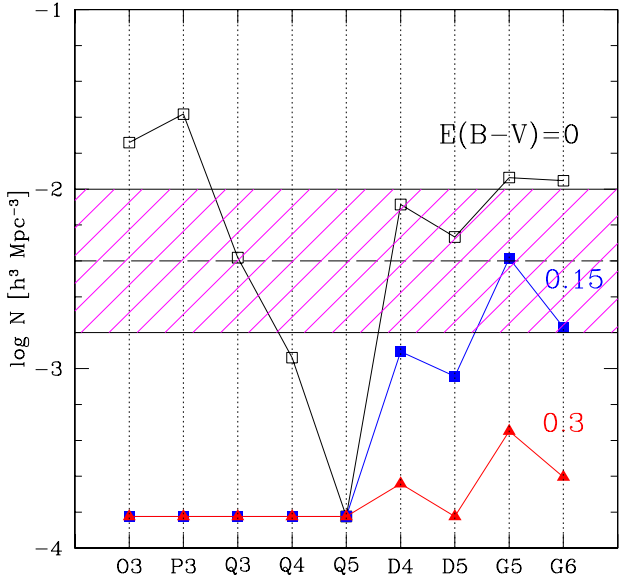,height=3in,width=3.2in, angle=0} 
\caption{Number density of galaxies for all the runs shown in 
Figure~\ref{colcol_Rcut_all.eps}.
Three different symbols denote three different values of 
extinction: $E(B-V)=0.0$ (black open square), 0.15 (blue filled square), 
and 0.3 (red filled triangle). Points with the same value of $E(B-V)$
are connected to guide the eye. A conservative range of the observed number 
density of LBGs is shown as the shaded region, with a central value of 
$4\times 10^{-3} {h}^3 \mpc^{-3}$ \citep{Ade03}. 
All the cases with zero LBGs are indicated as $N=-3.8$ dex for 
plotting purposes.}
\label{nden.eps}
\end{figure}

We now investigate the number density of LBGs in the simulation.  In
Figure~\ref{nden.eps}, we plot the number density of galaxies that
satisfy the colour-colour selection criteria of \citet{Ste03}, for all
the runs shown in Figure~\ref{colcol_Rcut_all.eps}.  Three different
symbols represent the three different values of extinction we used:
$E(B-V)=0.0$ (black open squares), 0.15 (blue filled squares), and 0.3
(red filled triangles). The points for the same value of $E(B-V)$ are
connected to guide the eye. 
All the cases with zero LBGs are indicated as $N=-3.8$ dex for 
plotting purposes.
A conservative range for the observed number densities of LBGs 
is shown as a shaded region, with a median value of $4\times 10^{-3}
{h}^3\mpc^{-3}$ \citep{Ade03}.  We note that \citet{Gia01} reported a
slightly smaller value of $2-3\times 10^{-3}{ h}^3\mpc^{-3}$.

From this figure, we find that the D4 \& D5 runs with $E(B-V)=0$
and the G5 \& G6 run with $E(B-V)=0.15$ appear to have reasonable 
number densities, while all runs with $E(B-V)=0.3$ seem to underpredict 
the number density. The O3 (no-wind) and P3 (weak wind) runs with $E(B-V)=0$ 
overpredict the number density significantly.
Apparently, without strong feedback by galactic winds and extinction, 
the LBGs simply become too bright and hence too abundant above a given
brightness limit. Other null results of the Q-series are affected by 
the small box size of the simulation.
We will show the effect of the box size more explicitly when we discuss 
the luminosity function of galaxies in Section~\ref{section:lf}.

\begin{figure}
\epsfig{file=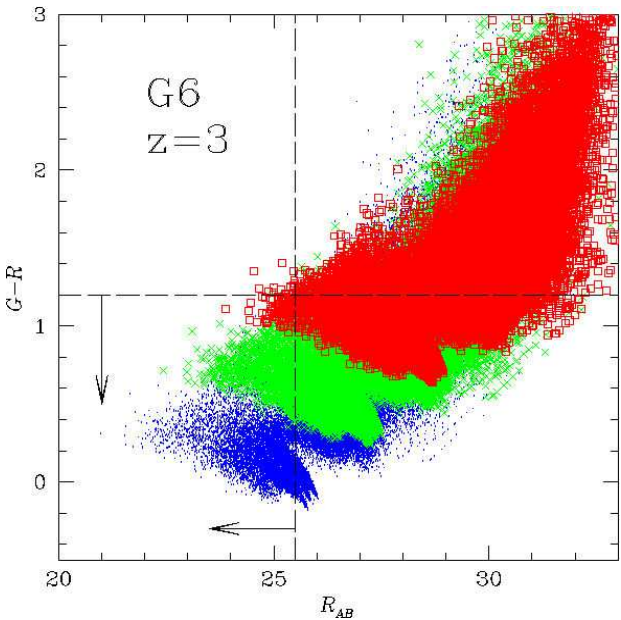,height=3.2in,width=3.2in, angle=0} 
\caption{Colour-magnitude diagram for the `G6'-run at $z=3$.
The distribution of galaxies is shown in the plane of $R$ 
apparent magnitude and $G-R$ colour. Three different symbols give results  
for three different values of extinction: $E(B-V)=0.0$ (blue dots), 
0.15 (green crosses), and 0.3 (red open squares). The long-dashed
lines and the arrow indicate the colour-selection criteria 
used by \citet{Ste03} to identify LBG candidates at $z\sim 3$.}
\label{Rmag_GRcol.eps}
\end{figure}


\section{Colour-magnitude relation and stellar masses}
\label{section:colour-magnitude}

In Figure~\ref{Rmag_GRcol.eps}, we show the distribution of galaxies
at $z=3$ in the `G6'-run on the plane of $R$-band apparent magnitude
and $G-R$ colour. We here chose the `G6'-run because it gives 
a reasonable fit to the observed luminosity functions and the number 
density, and it has higher resolution than the G5 run.  
Again, we use three different symbols for three different
values of extinction: $E(B-V)=0.0$ (blue dots), 0.15 (green crosses), and
0.3 (red open squares). The long-dashed lines and the arrows indicate
the colour-selection criteria applied by \citet{Ste03} to select LBG
candidates at $z\sim 3$.

We see that most of the galaxies brighter than $R=25.5$ automatically
satisfy the criterion $G-R<1.2$, and only a small fraction of galaxies
with $R<25.5$ fall out of the region.
There is a significant population of dim ($R>28$) galaxies with
$G-R>1.2$. We will see below that these are low-mass galaxies with 
stellar masses $M_{\star}\le 10^{10}\himsun$.

\begin{figure*}
\epsfig{file=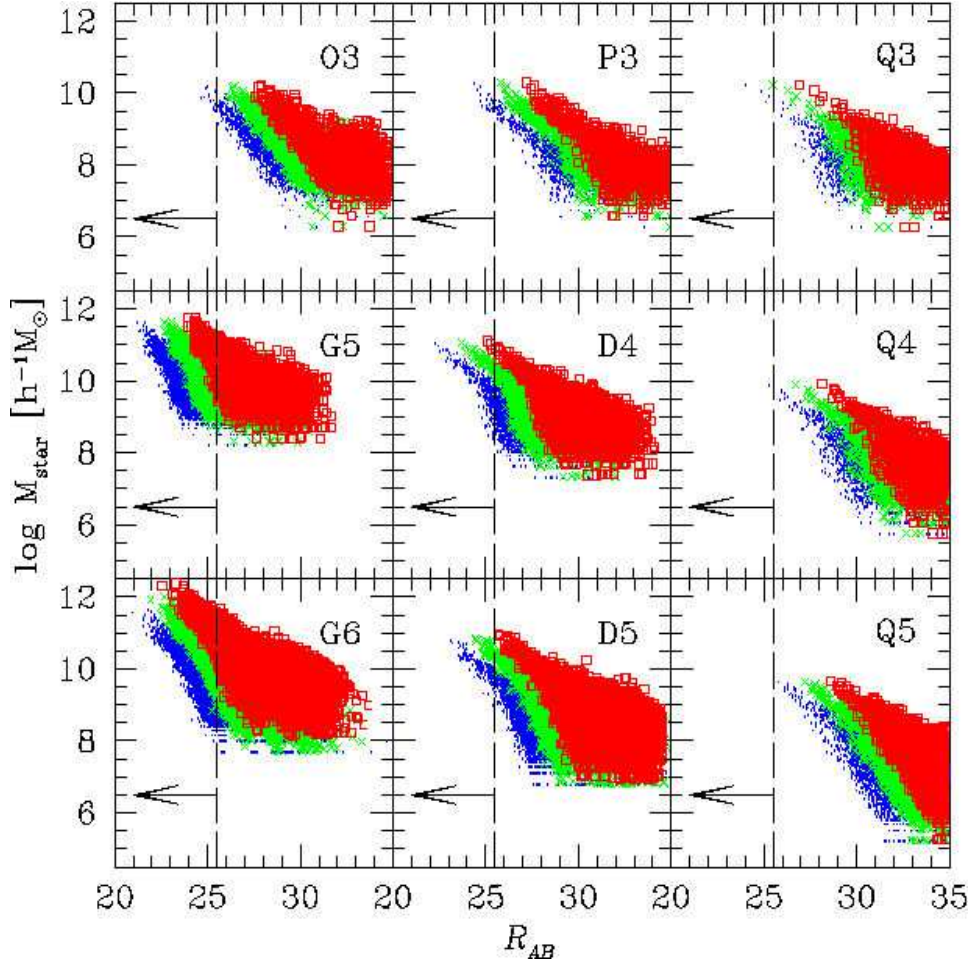,height=5in,width=5in, angle=0} 
\caption{Stellar mass vs. apparent magnitudes of simulated galaxies.
Three different symbols are used to show results for three different
values of extinction: $E(B-V)=0.0$ (blue dots), 0.15 (green crosses), and
0.3 (red open squares).  The vertical long-dashed line and the arrow
indicates the magnitude cut used by \citet{Ste03}.}
\label{Rmag_Mstar_all.eps}
\end{figure*}

In Figure~\ref{Rmag_Mstar_all.eps}, we show the $R$-band apparent
magnitude vs. stellar mass of simulated galaxies at $z=3$. As before,
we plot results for three values of extinction, using different
symbols. We also mark the magnitude limit $R<25.5$ used by
\citet{Ste03} with a vertical long-dashed line and arrows.

From this Figure, we see that the LBGs in the Q-series with $R<25.5$
have typically stellar masses in the range $M_{\star}=10^9 -
10^{10}\himsun$, while those in `D5' take a somewhat wider interval,
nearly covering the range $10^8 - 10^{11}\himsun$.  
The G6 run covers the even wider mass range 
$M_{\rm star}=10^8 - 10^{12}\himsun$
depending on the value of extinction. 
As the wind strength is increased from O3 to P3, and then to Q3-run, we observe
that the distribution of galaxies becomes slightly sparser at the most
luminous (i.e.~massive) end of the distribution.  We see the same
effect more clearly in the luminosity functions discussed in
Section~\ref{section:lf}. Note that since the numerical resolution is
identical for O3, P3, and Q3, the total baryonic mass in the simulation 
box is the same for these runs, but the runs with stronger winds convert 
less of their baryonic mass components into stars.

\begin{figure*}
\epsfig{file=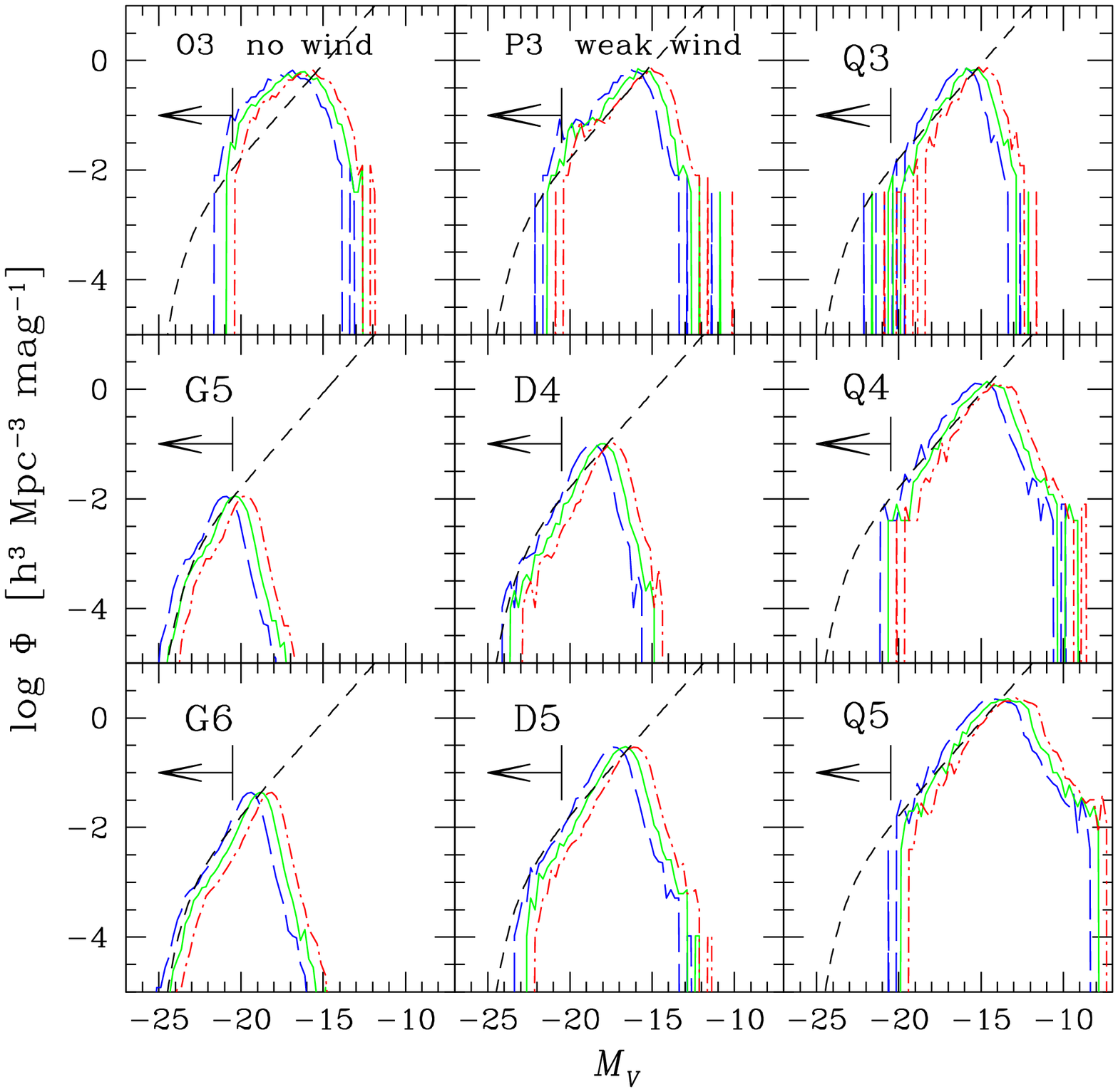,height=5in,width=5in, angle=0} 
\caption{Rest-frame V-band luminosity functions.
The three different lines are for three different values of extinction: 
$E(B-V)=0.0$ (blue long-dashed), 0.15 (green solid), and 0.3 (red 
dot-dashed). The black short-dashed curve is the observational estimate
of the rest-frame V-band luminosity function by \citet{Sha01}, and the
observational magnitude limit of $M_V= -20.5$ is shown by the arrows.
Note that the Schechter function is plotted down to the fainter magnitudes
only to guide the eye for comparison with the simulation results.}
\label{lf_V.eps}
\end{figure*}

In simulations with larger box size (D- and G-series), the masses of
the most luminous galaxies found are substantially larger compared
with those in the Q-series. This is primarily a result of a finite
box size effect; the brightest galaxies have a very low space-density,
such that they are simply not found in simulations with too small a
volume. On the other hand, the mass resolution of the small box size
simulations is much better, so that they can resolve a much larger
number of dim low-mass galaxies. Note that the lower cut-off of the
distribution in each panel is given by the stellar particle mass of
the corresponding simulation.

Interestingly, we find that the amount of stellar mass contained in
the LBGs that satisfy the colour selection criteria is in general far
from being close to the total stellar mass.  In the `D5'-run, for
example, the LBGs that pass the criteria contain only 20\% of all the 
stars for an extinction of $E(B-V)=0.0$.  In the `G5'-run,
the fraction is 57\% and 20\% for $E(B-V)=0.0$ and 0.15, respectively.
In the `G6'-run, the fraction is 52\% and 27\% for $E(B-V)=0.0$ and 
0.15, respectively. This means that the current surveys of LBGs 
fail to account for more than half of the stellar mass in the
Universe. \citet{Nachos} reach a similar conclusion from different 
theoretical arguments by comparing the results of two different 
types of hydrodynamic simulations and the theoretical model of 
\citet{Her03} with near-IR observations of galaxies 
\citep[e.g.][]{Dickinson, Rudnick03}. Such missed stellar masses 
might be hidden in the red population of galaxies as suggested by 
\citet{Franx}, but in our simulations the missed stellar mass fractions
are mainly contained in the faint galaxies below the observational 
magnitude limit.


\section{Luminosity Functions} 
\label{section:lf}

\subsection{Rest-frame V-band luminosity function}
\label{section:lf_V}

\begin{figure}
\epsfig{file=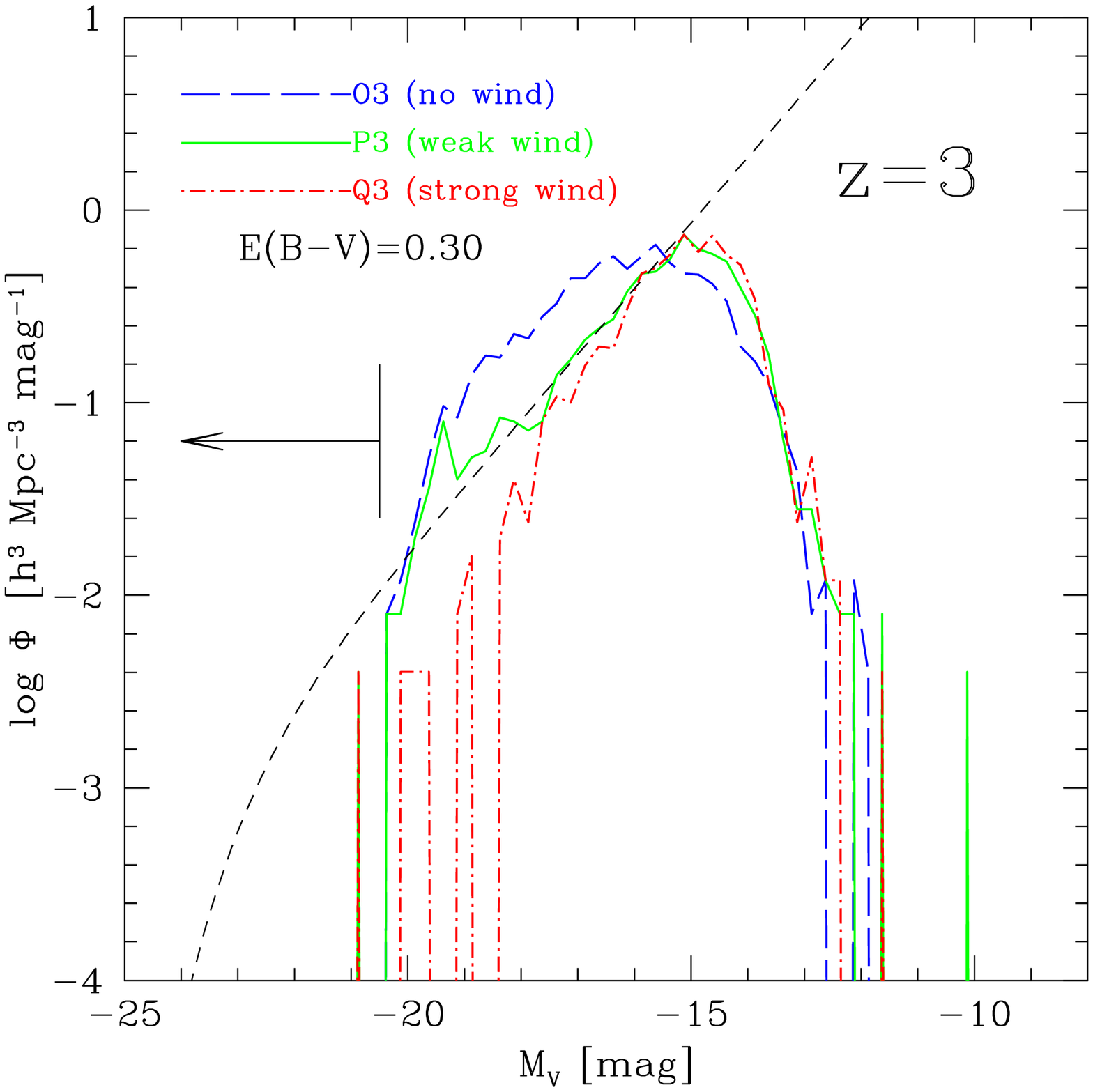,height=3.2in,width=3.2in, angle=0} 
\caption{Comparison of the rest-frame V-band luminosity functions of 
the O3 (no wind), P3 (weak wind), and Q3 (strong wind) simulations for 
$E(B-V)=0.3$. The black short-dashed curve is the observational estimate 
by \citet{Sha01}, and the observational magnitude limit of $M_V= -20.5$
is shown by the arrows. Note that the Schechter function is plotted 
down to the fainter magnitudes only to guide the eye for comparison 
with the simulation results.}
\label{lf_wind.eps}
\end{figure}

In Figure~\ref{lf_V.eps}, we show the rest-frame V-band luminosity
functions for all the simulated galaxies at $z=3$. 
We have also tried restricting the luminosity functions to only those 
galaxies that satisfy the colour-selection criteria. Most of the galaxies 
that fall out of the colour-selection criteria are the less luminous ones 
as demonstrated in Figure~\ref{Rmag_GRcol.eps}. This only affects the 
faint-end of the luminosity function not probed by current LBG surveys.
Therefore we show the luminosity functions for all galaxies in our 
simulations in Figures~\ref{lf_V.eps}, \ref{lf_wind.eps}, \& 
\ref{lf_Rmag_all.eps} for clarity. As before, we plot
results for three different values of extinction: $E(B-V)=0.0$ (blue
long-dashed), 0.15 (green solid), and 0.3 (red dot-dashed). Note that
stronger extinction simply shifts the luminosity function to the fainter
side, without changing its shape.  The black short-dashed line included 
in the figure gives the observational estimate of the rest-frame V-band
luminosity function by \citet{Sha01} with a faint-end slope of
$\alpha=-1.85$, normalisation $\Phi^*=0.18\times 10^{-2}
h^3\mpc^{-3}$, and characteristic magnitude $M*=-22.21+5\log h =
-22.98$, for $h=0.7$. The observational magnitude limit of $M_V= -20.5$
is shown by the arrows.

The most prominent feature seen in all panels is that the luminosity
functions of the simulated galaxies are all very steep, with a faint-end
slope comparable to $\alpha \sim -2$, which is the slope of the dark
matter halo mass function. This suggests that the strong feedback
included in the simulations has not been able to reduce the
luminosities of low-mass galaxies much more strongly than those of
more massive systems; if such a differential effect existed, it should
have manifested itself as a flattening of the faint-end compared to
the halo mass function. At face value, however, the observational data
actually support a rather steep faint-end slope at $z=3$, quite close
to that of the halo mass function.  One should note, however, that the
observational estimate of the slope $\alpha$ at $z=3$ is very
uncertain because the observations can only reach down to a magnitude
of $M_V\sim -20.5$, even with 8-meter class telescopes. 

When compared with the observational fit of \citet{Sha01}, it is clear
that the `Q'-series are deficient in the brightest galaxies at the high
luminosity-end of the luminosity function.  This can be understood as
a result of the small box size ($\Lbox = 10\,\himpc$) of these runs,
which do not have large enough volume to allow a faithful sampling of
rare, bright objects.  As the box size becomes larger from the
`Q'-series to the `D'-series ($\Lbox=33.75\,\himpc$), and then to the
`G'-series ($\Lbox=100\,\himpc$), this situation improves however. More
and more of the luminous objects can then be found, and the agreement
with the observation becomes better at the bright-end of the
luminosity function.

On the other extreme of the luminosity distribution, we see that
increasing the resolution from Q3 to Q4, and then to Q5, allows
inclusion of ever fainter objects, as expected. Therefore the
luminosity function becomes wider towards the fainter end. Note,
however, that the bright-end hardly changes, suggesting good
convergence in the simulation results for the massive galaxies.

The run `O3' (no wind run) slightly overpredicts the number of
galaxies compared to observations, arguing for the need of stronger
feedback. To investigate this further, we explicitly compare in
Figure~\ref{lf_wind.eps} the rest-frame V-band luminosity functions
as a function of wind strength.  We show the results for the simulations
O3 (no wind), P3 (weak wind), and Q3 (strong wind), using an extinction
of $E(B-V)=0.3$.  For reference, we include the observational
estimate by \citet{Sha01} as a black short-dashed curve.  Clearly,
stronger galactic winds reduce the number of bright galaxies, as a
result of the suppression of star formation they incur.
The same effect has been found in the Eulerian hydrodynamic simulations 
by \citet[][in preparation]{NCO}.


\subsection{Luminosity functions as observed at z=0}
\label{section:lf_R}

\begin{figure*}
\epsfig{file=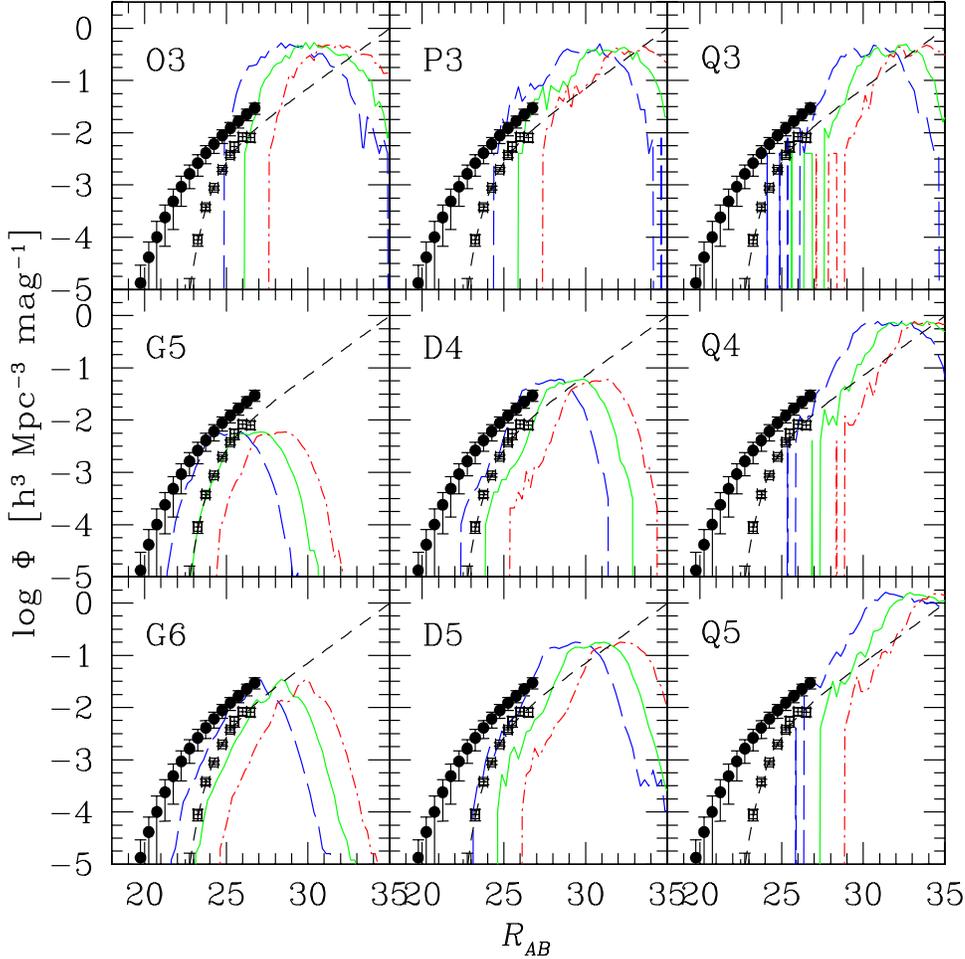,height=5in,width=5in, angle=0} 
\caption{R-band luminosity functions as observed at $z=0$.  The three
different lines are for three different values of extinction:
$E(B-V)=0.0$ (blue long-dashed), 0.15 (green solid), and 0.3 (red
dot-dashed). The open square symbols give the observed luminosity
function, while the solid circles are the dust corrected data points,
as shown in Fig.~13 of \citet{Ade00}.
\label{lf_Rmag_all.eps}}
\end{figure*}

In Figure~\ref{lf_Rmag_all.eps}, we show the $R$-band luminosity
function of simulated $z=3$ galaxies as seen at the present epoch.
As in Figure~\ref{lf_V.eps}, the three different lines are for three 
different values of extinction: $E(B-V)=0.0$ (blue long-dashed), 
0.15 (green solid), and 0.3 (red dot-dashed).  
The open square symbols give the direct observational result for 
the luminosity function of LBG galaxies in the observed frame,
while the solid circles are the data corrected for dust extinction, as
given in Fig.~13 of \citet{Ade00}. For the uncorrected data points
(open squares), a Schechter function fit with parameters $\Phi^* =
4.4\times 10^{-3}$ $h^3 \mpc^{-3} {\rm mag}^{-1}$, $M^*=24.54$, and
$\alpha=-1.57$ is also drawn as a short-dashed line. As discussed in
Section~\ref{section:method}, we applied the \citet{Calzetti}
extinction law locally at $z=3$ to the simulated galaxies, and then
redshifted their spectra, while simultaneously accounting for the
absorption by the IGM following the prescription of \citet{Madau95}.
The combined effects of redshifting and absorption by the IGM change
the shape of the luminosity function in a complex manner compared to
the rest-frame.

For the Q-series, the agreement between the simulations and the
observational data of \citet{Ade00} is not very good. On the bright
end, these runs simply lack galaxies due to their small box size,
explaining the discrepancy. But the well resolved faint-end appears to
show some excess over the observational result.  
The G-series agrees
better with the observations however. In particular, the G5 and G6 runs 
with $E(B-V)=0.15$ agree well with the uncorrected observed luminosity
function, and  the case for $E(B-V)=0.0$ is reasonably close to 
the dust corrected one of \citet{Ade00}.

\citet{Ade00} corrected their observed luminosity function for dust by 
adopting the relation $A_{1600}=4.43 + 1.99\beta$ \citep{Meurer} for the
extinction in magnitudes at 1600 \AA, where $\beta$ is the UV slope 
of the spectrum which has a distribution in the range of $[-3, 0]$ 
with a peak around $-1.5$, corresponding to $A_{1600}\sim 1.4$ mag 
\citep[see Fig. 12 of][]{Ade00}.
This peak value is roughly consistent with the value of $E(B-V)=0.15$
we adopted here with the \citet{Calzetti} extinction law $k(\lambda)$, 
because $E(B-V) k(1600\AA)\sim 1.5$ mag. Therefore the rough agreement
between the simulated luminosity function with $E(B-V)=0.0$ for 
G5 and G6-run and the dust corrected data points of \citet{Ade00} is 
encouraging, although not perfect.

Also, note that the peaks of the luminosity functions of the G5 run 
with $E(B-V)=0.0$ are on the brighter side of $R=25.5$ owing to the lower 
resolution.
This results in the increase of the number density of the LBGs that 
satisfy the colour-selection criteria when the resolution is increased
from the G5 to G6 run as described in Section~\ref{section:color-color}.


Overall, the comparison of the simulated luminosity functions with the 
observational results presented in Sections~\ref{section:lf_V} and 
\ref{section:lf_V} stresses the importance of having a large simulation 
box size ($> 100\himpc$) in order to obtain reasonable agreement 
at the bright end.
In our simulations, only the G-series contain a sufficient 
sample of LBGs brighter than $R=25.5$ with $E(B-V)=0.15$. 


\section{Star Formation History of LBGs}
\label{section:sf}

\begin{figure}
\epsfig{file=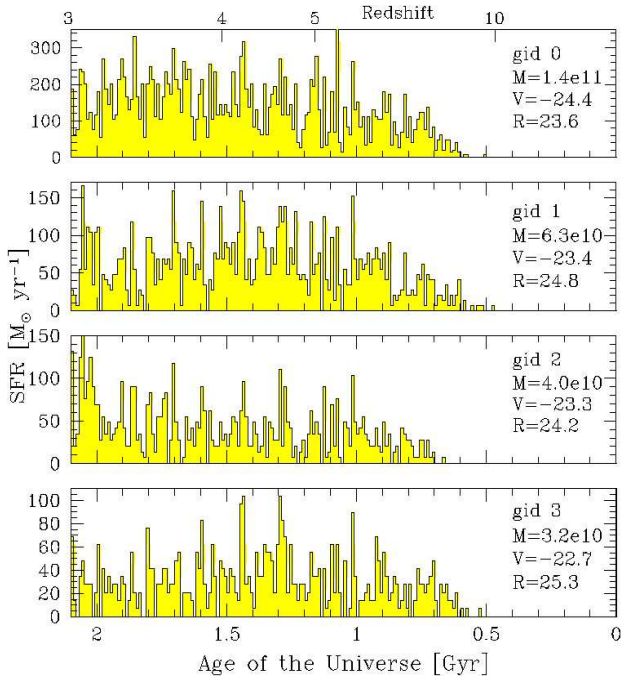,height=3.5in,width=3.2in, angle=0} 
\caption{Star formation histories of selected galaxies in the
`G6'-run with the bin-size of 10 Myrs.  Galaxy ID, stellar mass 
in units of $\himsun$, rest-frame $V$-band magnitude, and 
apparent $R$ magnitude for $E(B-V)=0.15$ are indicated in each
panel for the corresponding galaxy.
\label{sf_LBG.eps}}
\end{figure}

In Figure~\ref{sf_LBG.eps}, we show four examples of typical star
formation (SF) histories of galaxies in the `G6' run with
a bin-size of 10 Myrs, as derived from the age distribution of stars 
found in each galaxy including all the progenitors that merge prior 
to $z=3$.  We here chose the G6 run because it gives a reasonable 
agreement with observations both for the $R$-band and 
rest-frame $V$-band luminosity functions, and it has higher 
resolution than the G5 run.
On the right hand side of each panel, we indicate for each
galaxy an ID, its stellar mass in units of $\himsun$, its apparent
$R$ magnitude (for $E(B-V)=0.15$), and its rest-frame $V$-band 
magnitude.

A notable feature present in all the panels is the numerous spikes of 
starbursts lying on top of a relatively continuous component. 
The starbursts last $10-20$ Myrs, but the underlying smooth component 
shows only moderate variations in its rate when averaged over bins 
of 100 Myrs beginning from high redshift ($z\sim 10$) to $z=3$ 
in these galaxies. For example, the galaxy shown in the top panel, 
which is one of the most massive galaxies in the G6 run with a 
stellar mass of $\Mstar = 1.4\times 10^{11}\himsun$ and rest-frame 
V-band magnitude of $-24.4$, has continuously formed stars from 
$z=10$ to $z=3$ at a typical rate of $\sim 40~\Msun~\yr^{-1}$.

In the other panels, we show galaxies of progressively smaller total
mass, with correspondingly lower levels of star formation.  Note
that the galaxy on the bottom has a stellar mass of only $M_{\star} =
3.2\times 10^{10}\himsun$, yet it still satisfies the colour-colour
selection criteria, although its luminosity and star formation rate
are substantially smaller compared with the galaxy shown in the top
panel.  The recent star formation near $z=3$ has helped to bring the
brightness of this relatively small mass galaxy above the limit of
$R=25.5$.

Interestingly, the underlying continuous component of the star formation 
histories measured here are much smoother than the ones found by 
\citet{Nag02} when analysing a Eulerian hydrodynamic simulation. 
Their starbursts were more extended on the order of 100 Myrs, and 
more sporadic.  This difference is likely related to
the different models used by the codes for the treatment of the
physics of star formation. In the SPH methodology we investigate here,
a sub-resolution multiphase model for the ISM was used which has a
self-regulating property; i.e.~gas that cools smoothly onto the ISM is
also consumed in a smooth fashion by star formation, and only gas-rich
major mergers can trigger starbursts.


\section{Discussion}
\label{section:discussion}

We have used state-of-the-art hydrodynamic simulations of structure
formation to study the properties of LBGs in a $\Lam$CDM universe.
Our simulations use a new `entropy conserving' SPH formulation that
minimises systematic inaccuracies in simulations with cooling, as well
as an improved model for the treatment of the multiphase structure of
the ISM in the context of star formation and feedback \citep{SH03a}.
For the first time, our study uses a large series of simulations to
investigate the properties of LBGs, allowing an exploration of an 
unprecedented large dynamic range in both mass and spatial scales, 
while simultaneously providing reliable estimates of systematic 
effects due to numerical resolution.

By comparing our results for LBG properties in simulations with
different box size, we showed that a sufficiently large volume (at
least $\Lbox>50\,\himpc$) is crucial to faithfully sample the LBG
population. In particular, the bright-end of the LBG luminosity
function is invariably incomplete in simulations with small box size.
Of equal importance is a proper treatment of effective feedback
processes. By considering a series of runs with different strengths of
galactic winds, we showed that without an effective feedback process,
the number of LBGs is overpredicted. However, for our default model of
strong winds, a reasonable space density of bright LBGs is obtained.
In particular, we found that our G-series with a box size of
$\Lbox=100\,\himpc$ has a quite plausible LBG population, with
luminosity functions in both rest-frame $V$-band and observed-frame
$R$-band that match the observations reasonably well, at least at the 
bright-end.

Perhaps the most important conclusion of this paper is that the
observed properties of LBGs, including their number density, colours,
and luminosity functions, can be well explained if the LBGs are simply
associated with the most massive galaxies at $z=3$, with median
stellar mass of $M_* \sim 10^{10}\himsun$. This conclusion is
consistent with earlier numerical studies based on hydrodynamic
simulations of the $\Lam$CDM model \citep{Dave99, KHW99, Nag02, Wei02}
as well as some semi-analytic models \citep{Bau98, Kau99}, 
and does not provide direct support for alternative models
which suggest that LBGs are star-bursting low-mass systems that later
evolve into low-mass spheroids at $z=0$. This point is corroborated by
the high rates of star formation with $>10 ~\Msun~\yr^{-1}$ seen over
extended periods of time of order 1~Gyr in the simulated galaxies,
leading to the build-up of typical stellar masses of $10^{10}\himsun$ at
$z=3$. These comparatively steady star formation histories are also
consistent with observational studies by \citet{Papovich01} and
\citet{Sha01}.  Note however that these same observations also find
some evidence for starburst activity ($\ge 100\, \Msun \yr^{-1}$)
within the last 500 Myr before $z=3$, lasting for periods of $\sim
100$~Myr. Such long violent bursts are not seen in our SPH simulations, 
but they were present in the simulations analysed by \citet{Nag02}.
Instead, multiple shorter bursts with time-scale of $\sim 10-20$~Myr 
are seen in our SPH simulations.

A recent study by \citet{WW} however claims that the kinematic
measurements of \citet{Pettini01} and \citet{Erb} favour the
picture of LBGs being low-mass starbursting systems. But the measured
`rotation curves' are strongly dependent on seeing, and it is very
likely that these measurements underestimate the true rotation curves
(M. Pettini, private communication), casting some doubt on these
claims.  To settle this issue, we have to wait for the detection of a
definite flattening of the rotation curves, which might become
possible in the near future with near-IR spectrographs coupled to
adaptive optics systems.  Another difficulty with the low-mass
starbursting-galaxy picture for LBGs is that such systems would be
expected to have relatively low chemical abundances, whereas at least
the brighter LBGs have near-solar metallicity.

The luminosity functions we measure for our simulated LBG galaxies are
characterised by a steep faint end, with a slope close to $\alpha \sim
-2$.  This is the slope of the dark matter halo mass function, so in
our simulations, the galaxy luminosity function seems to largely
follow the halo mass function at the low-mass end.  Observationally,
we know that at $z=0$ the faint-end slope of galaxies is certainly
much shallower than this, with a slope of order $\alpha\sim -1.2$
\citep{Blanton}. At high-redshift, the faint-end of the galaxy
luminosity function is not nearly as well constrained, however, and
may be much steeper. In fact, \citet{Sha01} report a slope of
$\alpha=-1.85$ for LBG galaxies at $z=3$ which is still consistent
with our simulations.  However, as the authors explain, this should be
taken with some caution because their observations only reach an
absolute $V$-band magnitude of $-20.5$, which is not deep enough to
constrain the faint-end slope reliably. If confirmed, a steep
luminosity function at $z\simeq 3$ would in any case have to evolve
into a much flatter one at low redshift. Whether such a scenario is 
viable or not will be tested by current and future very deep surveys, 
such as DEEP2 \citep{Davis, Madge, Coil}. 

Barring confirmation of a rapid evolution of the faint-end slope, our
results therefore hint that the simulations still overproduce the
number of low-mass galaxies, despite the inclusion of strong feedback
processes that are capable of accounting for the observed brightness
of $L_{\star}$-galaxies. 
The moderating effects of feedback on star formation
do depend on galaxy mass in the physical model followed by the
simulations in the sense that, for a given amount of star formation, 
the feedback by winds does comparatively more damage for small mass 
galaxies. This is because towards smaller galaxy mass scales, the 
winds find it easier to escape from their confining galactic potential 
wells. The winds also entrain more gas in the outflow for lower 
circular velocities, so that the net baryonic loss becomes larger 
towards smaller mass scales, making the total mass-to-light 
ratio of smaller halos in principle bigger.  But, this variation of
the feedback efficiency with galaxy mass may not be strong enough 
in the present set of simulations to close the significant gap between 
the faint-end slopes of halo mass function and galaxy luminosity 
function.  

Interestingly, the studies by \citet{Chiu, Nag01b}, and \citet{Nag02}
based on Eulerian hydrodynamic simulations did find a flatter 
faint-end slope \citep[see also][]{Harford}. Presently it is unclear 
whether this was due to resolution limitations in the low-mass end of the halo 
mass function, or due to genuine physical effects of the feedback 
model implemented in these Eulerian simulations. More work in the 
future will be needed to settle this very interesting question, 
which is of tremendous importance for the theoretical framework of 
hierarchical galaxy formation in CDM universes.


\section*{Acknowledgements}

We thank Kurt Adelberger for providing us with the $\U$, $G$, $R$
filter response functions and the data points in
Figure~\ref{lf_Rmag_all.eps}, and Max Pettini for useful comments.  We
also thank Antonella Maselli for refereeing our paper and valuable
comments which improved the manuscript.  This work was supported in
part by NSF grants ACI 96-19019, AST 00-71019, AST 02-06299, and AST
03-07690, and NASA ATP grants NAG5-12140, NAG5-13292, and NAG5-13381.
The simulations were performed at the Center for Parallel
Astrophysical Computing at the Harvard-Smithsonian Center for
Astrophysics.


\bsp

\label{lastpage}


\begin{thebibliography}{}

\bibitem[\protect\citeauthoryear{Adelberger et al.}{2003}]{Ade03} 
        Adelberger K. L., Steidel C. C., Shapley A. E., Pettini M., 2003, 
        ApJ, 584, 45

\bibitem[\protect\citeauthoryear{Adelberger \& Steidel}{2000}]{Ade00} 
        Adelberger K. L. \& Steidel C. C., 2000, ApJ, 544, 218

\bibitem[\protect\citeauthoryear{Adelberger et al.}{1998}]{Ade98} 
        Adelberger K. L., Steidel C. C., Giavalisco M., Dickinson M., 
        Pettini M., Kellogg M., 1998, ApJ, 505, 18

\bibitem[\protect\citeauthoryear{Aguirre et al.}{2001a}]{Aguirre01a} 
        Aguirre A., Hernquist L., Schaye J., Weinberg D., Katz N.,
        Gardner J., 2001a, ApJ, 560, 599

\bibitem[\protect\citeauthoryear{Aguirre et al.}{2001b}]{Aguirre01b} 
        Aguirre A., Hernquist L., Schaye J., Katz N., Weinberg D.,
        Gardner J., 2001b, ApJ, 561, 521

\bibitem[\protect\citeauthoryear{Baugh et al.}{1998}]{Bau98} Baugh C. M.,
        Cole S., Frenk C. S., Lacey C. G., 1998, ApJ, 498, 504 

\bibitem[\protect\citeauthoryear{Becker et al.}{2001}]{Beck01} 
        Becker R. H., et al., 2001, AJ, 122, 2850

\bibitem[\protect\citeauthoryear{Blanton et al.}{2001}]{Blanton}
        Blanton M., et al., 2001, AJ, 121, 2358 

\bibitem[\protect\citeauthoryear{Bruscoli et al.}{2003}]{Bruscoli}
        Bruscoli M., Ferrara A.,  Marri S., Schneider R., Maselli A., Rollinde E., Aracil B., 2003, MNRAS, 343, L41

\bibitem[\protect\citeauthoryear{Bruzual \& Charlot}{1993}]{BClib}
        Bruzual, G. A. \& Charlot, S., 1993, ApJ, 405, 538

\bibitem[\protect\citeauthoryear{Calzetti et al.}{2000}]{Calzetti}
        Calzetti, D., et al., 2000, ApJ, 533, 682

\bibitem[\protect\citeauthoryear{Chiu, Gnedin \& Ostriker}{2001}]{Chiu} 
        Chiu W. A., Gnedin N. Y., Ostriker J. P., 2001, ApJ, 563, 21 

\bibitem[\protect\citeauthoryear{Coil et al.}{2003}]{Coil} Coil, A., et al.,
        2004, ApJ, 609, 525

\bibitem[\protect\citeauthoryear{Croft et al.}{2001}]{Croft01} Croft R. A. C.,
        Di Matteo T., Dav\'e R., Hernquist L., Katz N., Fardal M. A.,
        Weinberg D.H., 2001, ApJ, 557, 67

\bibitem[\protect\citeauthoryear{Croft et al.}{2002}]{Croft02} Croft R. A. C.,
        Hernquist L., Springel V., Westover M., White M., 2002, ApJ, 580, 634

\bibitem[\protect\citeauthoryear{Dav\'e et al.}{1999}]{Dave99} Dav\'e R., 
        Hernquist L., Katz N., Weinberg D. H., 1999, ApJ, 511, 521

\bibitem[\protect\citeauthoryear{Davis et al.}{2002}]{Davis} 
        Davis, M., et al., 2002, SPIE (astro-ph/0209419)

\bibitem[\protect\citeauthoryear{Desjacques et al.}{2003}]{Des}
        Desjacques V., Nusser A., Haehnelt M. G., Stoehr F., 2004, 
        MNRAS, 350, 879

\bibitem[Dickinson et al.(2003)]{Dickinson} Dickinson M., Papovich C., 
        Ferguson H., Tam\'{a}s Budav\'{a}ri, ApJ, 2003, 587, 25

\bibitem[Erb et al.(2003)]{Erb} Erb D. K., et al., 2003, ApJ, 591, 110

\bibitem[Franx et al.(2003)]{Franx} Franx M., et al., 2003, ApJ, 587, L79

\bibitem[Giavalisco \& Dickinson(2001)]{Gia01} Giavalisco, M. \& 
        Dickinson, M., 2001, ApJ, 550, 177

\bibitem[Giavalisco~\etal(1998)]{Giavalisco98} Giavalisco, M., Steidel,
                      C. C., Adelberger, K. L., Dickinson, M. E.,
                      Pettini, M., \& Kellogg, M. 1998, ApJ, 503, 543

\bibitem[\protect\citeauthoryear{Haardt \& Madau}{1996}]{Haa96} Haardt F., 
        Madau P., 1996, ApJ, 461, 20 

\bibitem[{Harford \& Gnedin(2003)}]{Harford} Harford A. G. \& 
        Gnedin N. Y., 2003, ApJ, 597, 74

\bibitem[\protect\citeauthoryear{Hernquist}{1993}]{Hern93} 
        Hernquist L., 1993, ApJ, 404, 717

\bibitem[\protect\citeauthoryear{Hernquist \& Springel}{2003}]{Her03} 
        Hernquist L., Springel V., 2003, MNRAS, 341, 1253

\bibitem[\protect\citeauthoryear{Jing \& Suto}{1998}]{Jing} 
        Jing Y. P., Suto Y., 1998, ApJ, 494, L5

\bibitem[\protect\citeauthoryear{Katz, Hernquist \& Weinberg}{1999}]{KHW99} 
        Katz N., Hernquist L., Weinberg D. H., 1999, ApJ, 523, 463

\bibitem[\protect\citeauthoryear{Katz et al.}{1996}]{Katz96} Katz N., 
        Weinberg D. H., Hernquist L., 1996, ApJS, 105, 19 

\bibitem[\protect\citeauthoryear{Kauffmann et al.}{1999}]{Kau99} 
        Kauffmann G. A. M., Colberg J. M., Diaferio A., White S. D. M., 1999,
        MNRAS, 307, 529

\bibitem[\protect\citeauthoryear{Kollmeier et al.}{2003}]{Kol03} 
        Kollmeier J. A., Weinberg D. H., Dav\'e R., Katz N., 2003, ApJ, 594, 75

\bibitem[Lowenthal~\etal(1997)]{Lowenthal97} Lowenthal, J. D., et al., 1997, ApJ, 481, 673

\bibitem[\protect\citeauthoryear{Madau}{1995}]{Madau95} Madau, P., 1995, ApJ, 441, 18

\bibitem[\protect\citeauthoryear{Madgewick et al.}{2003}]{Madge} Madgewick, D., et al. 2003, MNRAS, 344, 847
  
\bibitem[\protect\citeauthoryear{Maselli et al.}{2003}]{Maselli} Maselli A., Ferrara A., Bruscoli M., Marri S., Schneider R., MNRAS, 2004, 350, L21

\bibitem[\protect\citeauthoryear{Meurer, Heckman, \& Calzetti}{1999}]{Meurer}
        Meurer G. R., Heckman T. M., Calzett D., 1999, ApJ, 521, 64 

\bibitem[\protect\citeauthoryear{Mo \& Fukugita}{1996}]{Mo96} Mo H. J., Fukugita, 1996, ApJ, 467, L9

\bibitem[\protect\citeauthoryear{Mo, Mao, \& White}{1999}]{Mo99} Mo H. J., 
        Mao S., White S. D. M., 1999, MNRAS, 304, 175

\bibitem[\protect\citeauthoryear{Nagamine et al.}{2001}]{Nag01b} Nagamine K.,
        Fukugita M., Cen R., Ostriker J. P., 2001, MNRAS, 327, L10

\bibitem[\protect\citeauthoryear{Nagamine}{2002}]{Nag02} Nagamine K., 
        2002, ApJ, 564, 73

\bibitem[\protect\citeauthoryear{Nagamine, Springel, \& Hernquist}{2004a}]{Nag03a} Nagamine K., Springel V., Hernquist L. 2004a, MNRAS, 348, 421 

\bibitem[\protect\citeauthoryear{Nagamine, Springel, \& Hernquist}{2004b}]{Nag03b} Nagamine K., Springel V., Hernquist L. 2004b, MNRAS, 348, 435

\bibitem[\protect\citeauthoryear{Nagamine et al.}{2004}]{Nachos} Nagamine K., Cen R., Hernquist L., Ostriker J. P., Springel V., 2004, ApJ, 610, 45

\bibitem[\protect\citeauthoryear{Nagamine, Cen, \& Ostriker}{2004}]{NCO} Nagamine K., Cen R., Ostriker J. P., 2004, in preparation

\bibitem[Papovich et al.(2001)Papovich, Dickinson, \& Ferguson]{Papovich01} 
        Papovich C., Dickinson M., \& Ferguson H. C., 2001, ApJ, 559, 620

\bibitem[\protect\citeauthoryear{Pearce et al.}{1999}]{Pearce99} 
        Pearce et al., 1999, ApJ, 521, 99

\bibitem[\protect\citeauthoryear{Pettini et al.}{2002}]{Pet02} Pettini M., 
        Rix S. A., Steidel C. C., Adelberger K. L., Hunt M. P., Shapley A. E.
        2002, ApJ, 569, 742 

\bibitem[Pettini~\etal(2001)]{Pettini01} Pettini M., ~\etal ~2001, 
        ApJ, 554, 981

\bibitem[Rudnick~\etal(2001)]{Rudnick01} Rudnick G., ~\etal ~2001, 
        AJ, 122, 2205

\bibitem[Rudnick~\etal(2003)]{Rudnick03} Rudnick G., ~\etal ~2003, 
        ApJ, 599, 847

\bibitem[Salpeter(1995)]{Salpeter} Salpeter E. E. 1955, ApJ, 121, 161

\bibitem[Sawicki \& Yee(1998)]{Sawicki98} Sawicki M. \& Yee H. K. C., 1998, 
        AJ, 115, 1329

\bibitem[\protect\citeauthoryear{Shapley et al.}{2001}]{Sha01} Shapley A. E., 
        Steidel C. C., Adelberger K. L., Dickinson M., Giavalisco M., Pettini M.
        2001, ApJ, 562, 95 

\bibitem[\protect\citeauthoryear{Sokasian et al.}{2003}]{Sok03}
        Sokasian A., Abel T., Hernquist L., Springel V., 2003, MNRAS, 344, 607

\bibitem[Somerville et al.(2001)Somerville, Primack, \& Faber]{Som01} 
        Somerville, R. S., Primack, J. R., \& Faber, S. M., 2001, MNRAS, 320, 504

\bibitem[\protect\citeauthoryear{Springel et al.}{2001}]{Spr01} 
        Springel V., White S.~D.~M., Tormen G., Kauffmann G., 2001, MNRAS, 328, 726
        
\bibitem[\protect\citeauthoryear{Springel, Yoshida \& White}{2001}]{Gadget} 
        Springel V., Yoshida N., White S.~D.~M., 2001, New Astronomy, 6, 79

\bibitem[\protect\citeauthoryear{Springel \& Hernquist}{2002}]{SH02} 
        Springel V., Hernquist L., 2002, MNRAS, 333, 649

\bibitem[\protect\citeauthoryear{Springel \& Hernquist}{2003a}]{SH03a} 
        Springel V., Hernquist L., 2003a, MNRAS, 339, 289

\bibitem[\protect\citeauthoryear{Springel \& Hernquist}{2003b}]{SH03b} 
        Springel V., Hernquist L., 2003b, MNRAS, 339, 312

\bibitem[\protect\citeauthoryear{Steidel et al.}{2003}]{Ste03} Steidel C. C., 
        Adelberger K. L., Shapley, A., \& Pettini, M., 2003, ApJ, 592, 728

\bibitem[Steidel~\etal(1998)]{Steidel98} Steidel C. C., Adelberger K. L., Dickinson M., Giavalisco M., Pettini M., Kellogg M. 1998, ApJ, 492, 428 

\bibitem[Steidel, Pettini, \& Hamilton(1995)]{Steidel95} Steidel C. C., Pettini M., \& Hamilton D., 1995, AJ, 110, 2519

\bibitem[Steidel \& Hamilton(1993)]{Steidel93} Steidel C. C. \& Hamilton D., 1993, AJ, 105, 2017

\bibitem[\protect\citeauthoryear{Weatherley \& Warren}{2003}]{WW}
        Weatherley S. J. \& Warren S. J., 2003, MNRAS, 345, L29 

\bibitem[\protect\citeauthoryear{Weinberg, Hernquist \& Katz}{2002}]{Wei02} 
        Weinberg D. H., Hernquist L., Katz N., 2002, ApJ, 571, 15

\bibitem[\protect\citeauthoryear{Yoshida et al.}{2002}]{Yoshida02} Yoshida N, 
        Stoehr F., Springel V., White S. D. M., 2002, MNRAS, 335, 762

\end{thebibliography}
\end{document}